\documentclass[preprint, sort&compress]{elsarticle}

\usepackage{graphicx}
\usepackage{times}
\usepackage[utf8]{inputenc}
\usepackage{url}
\usepackage{amssymb}
\usepackage{amstext}
\usepackage{amsmath}
\usepackage{booktabs}
\usepackage{tabularx}
\usepackage{threeparttable}
\usepackage{multirow, makecell} 
\usepackage{rotating}
\usepackage{nicefrac} 
\usepackage{setspace}
\usepackage[english]{babel}

\usepackage{hyperref}
\usepackage{cleveref}

\addto\extrasenglish{%
}

\usepackage{hyperref}
\addto\extrasenglish{%
}

\hypersetup{colorlinks=true, linkcolor=blue, anchorcolor=red, citecolor=blue, filecolor =red, urlcolor=red, pdfauthor=author}

\usepackage{pifont}

\usepackage{color, colortbl}

\usepackage{pgfplots}
\usepackage{layouts}
\usepackage{comment}
\usepackage{threeparttable}
\usepackage{bm}
\usepackage{enumitem}
\usepackage{setspace}

\usepackage{chngcntr}
\counterwithin{table}{section} 

\usepackage{times}
\usepackage{balance}  
\usepackage{soul}               
\usepackage{url}
\usepackage{textcomp}
\usepackage{mathrsfs}

\usepackage[font=scriptsize]{subfig}

\usepackage{dcolumn}
\usepackage{color}
\usepackage{array}

\usepackage{textcomp}
\usepackage{siunitx}

\definecolor{Gray}{gray}{0.9}

\usepackage{longtable}
\usepackage{pdflscape}

\pgfplotsset{compat=1.18}

\usepackage{lineno}

\DeclareSIUnit{\day}{\text{\ensuremath{\mathrm{day}}}}
\begin{document}

\begin{frontmatter}

\title{A Deep Learning Approach for Virtual Contrast Enhancement in Contrast Enhanced Spectral Mammography}

\journal{}
\author[aff1]{Aurora Rofena\corref{cor}\fnref{contrib}}
\ead{aurora.rofena@unicampus.it}

\author[aff1]{Valerio Guarrasi\fnref{contrib}}

\author[aff3]{Marina Sarli}

\author[aff3]{Claudia Lucia Piccolo}

\author[aff3]{Matteo Sammarra}

\author[aff3,aff4]{Bruno Beomonte Zobel}

\author[aff1,aff5]{Paolo Soda}

\address[aff1]{Unit of Computer Systems \& Bioinformatics, Department of Engineering \\ University Campus Bio-Medico, Rome, Italy}

\address[aff3]{Department of Radiology, Fondazione Policlinico Campus Bio-Medico, Rome, Italy}

\address[aff4]{Department of Radiology, University Campus Bio-Medico, Rome, Italy}

\address[aff5]{Department of Radiation Sciences, Radiation Physics, Biomedical Engineering, Umeå University, Sweden}

\cortext[cor]{Correspondence: Aurora Rofena, Paolo Soda}
\fntext[contrib]{These authors contributed equally to this work.}
\begin{abstract}

Contrast Enhanced Spectral Mammography (CESM) is a dual-energy mammographic imaging technique that first needs intravenously administration of an iodinated contrast medium; then, it  collects both a low-energy image, comparable to standard mammography, and a high-energy image. 
The two scans are then combined to get a recombined image showing contrast enhancement.
Despite CESM diagnostic advantages for breast cancer diagnosis, the use of contrast medium can cause side effects, and CESM  also beams patients with a higher radiation dose compared to standard mammography.
To address these limitations this work proposes to use deep generative models for virtual contrast enhancement on CESM, aiming to make the CESM  contrast-free as well as to reduce the radiation dose. 
Our deep networks, consisting of an autoencoder and two Generative Adversarial Networks, the Pix2Pix, and the CycleGAN, generate synthetic recombined images solely from low-energy images.
We perform an extensive  quantitative and qualitative analysis of the model's performance, also exploiting radiologists’ assessments, on a novel CESM dataset that includes 1138 images that, as a further contribution of this work, we make publicly available.
The results show that CycleGAN is the most promising deep network to generate synthetic recombined images, highlighting the potential of artificial intelligence techniques for  virtual contrast enhancement in this field.

\end{abstract}


\end{frontmatter}


\tikzset{every picture/.style={}} 

\section{Introduction} \label{sec:introduction}

Contrast Enhanced Spectral Mammography (CESM) is a dual-energy breast imaging technique that belongs to level II breast diagnostic.
Unlike standard mammography, also referred to as full-field digital mammography (FFDM), CESM involves the injection of an iodinated contrast medium that diffuses into the tumor tissue, enhancing lesion visibility~\cite{stateart}.
This results in improved diagnostic accuracy, especially in patients with heterogeneously dense and extremely dense breast tissue, which in FFDM may obscure small masses and reduce sensitivity~\cite{cesm}.
Furthermore, CESM is a valid alternative to contrast-enhanced breast magnetic resonance imaging (MRI), being more specific, accessible, cheaper, and preferred by patients~\cite{cesm}.
During CESM, both breasts are alternately compressed in craniocaudal and mediolateral oblique projections for approximately two minutes after contrast medium injection.
A low-energy (LE) and high-energy (HE) image are acquired in quick succession for each compression, and within 10 minutes from the start of contrast administration, the exam is completed.
The LE image is acquired below the k-edge of iodine (which is 33.2 keV), thus it is blind to iodine and comparable to FFDM, even though the iodinated contrast medium is already present within the breast. 
The HE images is acquired above the k-edge of iodine, and it is not suitable for diagnostic purposes. However, through logarithmic subtraction, it is combined in post-processing with the LE image to obtain the recombined image, also referred to as dual-energy subtracted (DES) image.
The DES image suppresses parenchymal tissue, leaving only areas of contrast enhancement visible.
Thus, CESM provides the radiologist with LE and DES image pairs for diagnosis~\cite{stateart}.

However, two factors may limit the adoption of CESM as a widespread screening technique.
First, although the administration of iodinated contrast agent is essential for the performance of CESM and is generally considerated safe, adverse effects may occur, such as hypersensitivity reactions and the possibility of contrast-induced nepropathy (CIN)~\cite{sideeffects}. 
The estimated incidence of hypersensitivity reactions ranges from 0.7\% to 3.1\%, with severe reactions occurring in only 0.02\% to 0.04\% of administrations~\cite{review}.
CIN is defined as a decline in renal function within three days of iodine-based contrast medium administration.
Its pathophysiology remains poorly understood, but pre-existing renal dysfunction appears to contribute~\cite{frasecin}.
Second, CESM requires a double energy exposure for each projection, i.e., it collects both the LE and HE images, resulting in a higher radiation dose compared to FFDM~\cite{review}.

To overcome these limitations, reducing the administration of the contrast medium and the amount of radiation by using Artificial Intelligence (AI) techniques is a viable solution that can be now investigated, thanks to the recent progress in Deep Learning (DL) and generative approaches, in particular.
Indeed, in recent years, the use of deep neural networks has become dominant in image-to-image translation for medical images~\cite{overview}.
Image-to-image translation involves converting images from the source domain $X$ to the target domain $Y$.
This approach is implemented in medical image analysis, including task such as noise reduction~\cite{denoising}, super-resolution~\cite{superresolution}, cross‑modality synthesis~\cite{ctfrommri}, and virtual contrast enhancement (VCE), which is the goal of our work.
Previous studies have explored the application of deep neural networks in VCE for MRI and computed tomography (CT) images.
The aim was to generate synthetic contrast-enhanced images from non-contrast images, with the goal of evaluating the feasibility of reducing contrast media usage while preserving image quality.
In this context, Kleesiek et al.~\cite{vcemri} trained a fully convolutional autoencoder with dropout to predict contrast enhancement from non-contrast multiparametric brain MRI scans, with 10 channels used for the model input.
Choi er al. (2021)~\cite{ct-vce-pix2pix} evaluated the 3D implementation of Pix2Pix for generating synthetic contrast-enhanced chest CT from non-contrast chest CT.
Xie et al.~\cite{res-cyclegan} exploited a cycle-consistent generative adversarial network (CycleGAN) framework to learn a mapping from non-contrast CT to contrast-enhanced CT, using a residual U-Net as the generator of the CycleGAN. 

\begin{figure*}[h!] 
    \includegraphics[width=\textwidth]{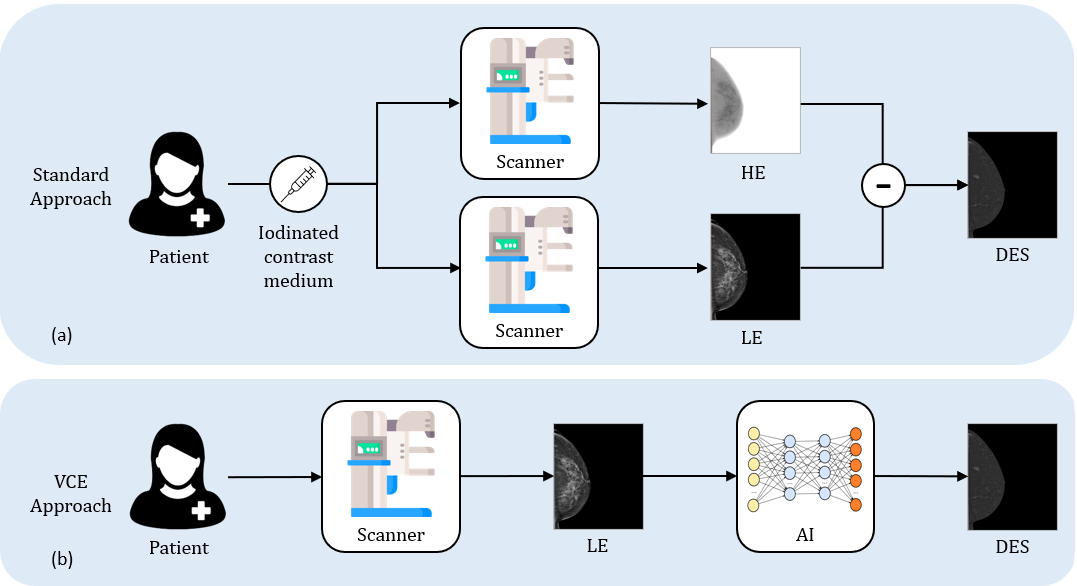}
    \caption{Illustration of DES images generation. (a) Standard approach with DES image obtained by injecting iodinated contrast medium into the patient and acquiring both the LE and HE images. (b) VCE approach proposed by our work, with DES images generated directly from the LE images, without the need to inject iodinated contrast medium or acquire HE images. }
    \label{fig:intro}
\end{figure*}

However, no one has yet analyzed VCE within the context of CESM. Nonetheless, we firmly believe that investigating this approach is a valid solution to overcome the use of contrast media and reduce exposure to dual energy.
Therefore, as depicted in \figurename~\ref{fig:intro}, our goal is to train and validate DL models to test if DES images can be synthesized directly from the corresponding LE images. By treating the LE images as FFDM images, as supported in~\cite{le-ffdm}, this approach would eliminate the necessity of administering contrast medium and acquiring the HE image, thus avoiding the need for a dual-energy exposure.

As part of our efforts to advance research in the CESM field, we release the dataset collected and used in this study.
This marks a significant milestone, as it is the first publicly available CESM dataset in DICOM format, with data anonymized to ensure privacy.
The dataset consists of 1138 images in DICOM format, comprising 569 LE images and 569 DES images, collected from 105 patients.
All relevant non-sensitive DICOM tags are accessible, and the dataset includes information extracted from medical reports and biopsy records.

The rest of the manuscript is organized as follows: Section~\ref{sec:materials} outlines the publicly released dataset used for the study.
Section~\ref{sec:methods} proposes the methods used, including neural network architectures, image pre-processing, and the training and evaluation process.
Section~\ref{sec:res and dis} presents and discusses the results obtained, while Section~\ref{sec:conclusion} offers our conclusions.

\section{Materials} \label{sec:materials}
In this work we collected and utilized an in-house dataset, named as CESM@UCBM in the following, which we make publicly available\footnote{http://www.cosbi-lab.it/cesmucbm/} after anonymizing sensitive data. 
The images are stored  in DICOM format, thus including imaging parameters, acquisition details, and annotations, which are useful for the analyses.

\tablename~\ref{tabdataset} summarizes the main characteristics of the  dataset.
It  consists of 1138 images, divided into 569 LE images and 569 DES images.
These images were obtained from 105 patients aged between 31 and 90 years, with an average age of 57.4 years and a standard deviation of 11.5 years.
They were enrolled in the study under the Ethical Committee approval PAR 51.23 OSS.
All the exams were performed at the Fondazione Policlinico Universitario Campus Bio-Medico in Rome between September 2021 and October 2022, using the GE Healthcare Senographe Pristina full-field digital mammography.
The resolution of 984 images is $2850 \times 2396$ pixels, while the remaining 154 images have a resolution of $2294 \times 1916$ pixels. 
Among the 1138 images in the CESM@UCBM dataset:
\begin{itemize}[noitemsep,nolistsep]
    \item 284 images show craniocaudal projection of right breasts, with 142 LE images and 142 DES images;
    \item 282 images show craniocaudal projection of left breasts, with 141 LE images and 141 DES images;
    \item 292 images show mediolateral oblique projection of right breasts, with 146 LE images and 146 DES images;
    \item 280 images show mediolateral oblique projection of left breasts, with 140 LE images and 140 DES images.
\end{itemize}

For 86 patients late-phase acquisitions are available, implying that the image of the same breast, in the same projection, was acquired multiple times within a few minutes.
Unilateral acquisitions were performed for 7 patients.

\begin{figure*}[h!] 
\centering
    \includegraphics[width=8 cm]{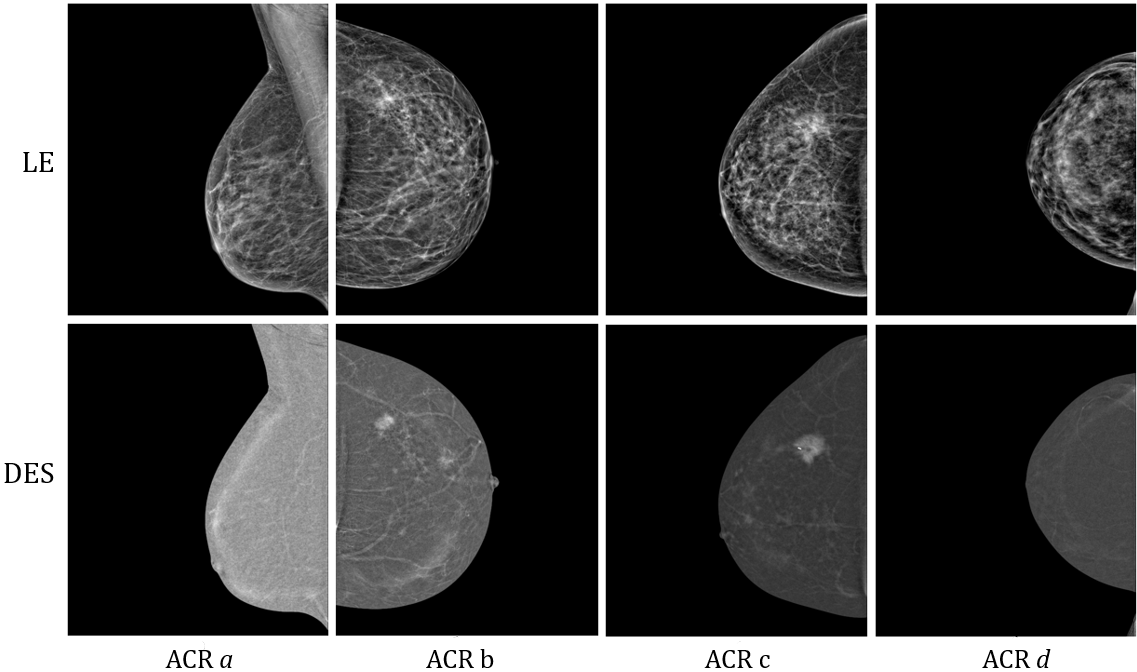}
    \caption{From left to right, pairs of LE (above) and DES images (below) from the CESM@UCBM dataset with ACR categories \textit{a}, \textit{b}, \textit{c}, \textit{d}.}
    \label{fig:acr}
\end{figure*}

Furthermore, for each patient, we extract information from the medical report and the biopsy, if available.
This includes the classification of breast composition with one of the 4 categories (\textit{a}, \textit{b}, \textit{c}, \textit{d}) defined by the fifth edition of the Breast Imaging Reporting and Data System (BI-RADS) released by the American College of Radiology (ACR)~\cite{acrbirads}.
The category \textit{a} indicates almost entirely fatty breasts and characterizes 130 images, of which 35\% show contrast enhancement.
The category \textit{b} identifies breasts with scattered areas of fibroglandular density and characterizes 360 images, of which 33\% show contrast enhancement.
The category \textit{c} indicates heterogeneously dense breast tissue and characterizes 414 images, of which 39\% show contrast enhancement.
Finally, the category \textit{d} identifies extremely dense breasts and characterizes 190 images, of which 36\% show contrast enhancement.
Instead, for 44 images the ACR category is not reported.
\figurename~\ref{fig:acr} compares 4 pairs of LE and DES images belonging to the CESM@UCBM dataset, which differ in the ACR category.

\begin{table*}[h!] 
\begin{center}
\resizebox{0.7\textwidth}{!}{
\begin{tabular}{llc}
\toprule
\multirow{4}{*}{\textbf{General Info}} & Period & September 2021 - October 2022 \\  & Source & Fondazione Policlinico UCBM\\ 
 & Patients & 105 \\ 
 & Total images & 1138 \\ 
 \midrule
\multirow{2}{*}{\textbf{Image Type}} & LE & 569 images \\ 
 & DES & 569 images \\ 
 \midrule
\multirow{2}{*}{\textbf{Image View}} & Craniocaudal & 566 images \\ 
 & mediolateral oblique & 572 images \\ 
 \midrule
\multirow{2}{*}{\textbf{Image Resolution}} & $2850 \times 2396$ & 984 images \\ 
 & $2294 \times 1916$ & 154 images \\ 
 \midrule
\multirow{4}{*}{\textbf{Patients' Age}} & $<$ 50 years & 31 patients \\ 
 & 50-59 years & 37 patients \\ 
 & 60-69 years & 19 patients \\ 
 & $\geq$ 70 years & 18 patients \\ 
 \midrule
\multirow{3}{*}{\textbf{Biopsy}} & Malignant & 258 images \\ 
 & Benignant & 104 images \\ 
 & Borderline & 16 images \\ 
 \midrule
\multirow{5}{*}{\textbf{ACR category}} & \textit{a} & 130 images \\ 
 & \textit{b} & 360 images \\ 
 & \textit{c} & 414 images \\ 
 & \textit{d} & 190 images \\ 
 & not reported & 44 images \\ 
 \midrule
 \multirow{1}{*}{\textbf{BI-RADS descriptors}} & 1 - 6 & 31 patients \\ 
 \bottomrule
\end{tabular}}
\caption{Characteristics of the CESM@UCBM dataset.}
\label{tabdataset}
\end{center}
\end{table*}

In addition, for 31 patients, the reports are completed with the BI-RADS descriptors (from 1 to 6) associated with the probability of malignancy of the lesion and the actions to be taken to conclude the diagnostic-therapeutic management process.
Based on biopsies, in 258 images, 129 of which were LE and 129 were DES, malignant lesions were identified. In 104 images, 52 LE and 52 DES, benign lesions were identified. In 16 images, 8 LE and 8 DES, borderline lesions were identified. The remaining images did not show any tumor-related abnormalities.

\section{Methods} \label{sec:methods}
We design the VCE task on CESM images as an image-to-image translation problem, where the source domain $X$ consists of LE images, and the target domain $Y$ consists of DES images.
Considering $x$ as a generic LE image from $X$, and $y$ as the corresponding DES image from $Y$, we implement DL models $G$ that take $x$ as input and attempt to generate the corresponding virtual contrast-enhanced DES images, i.e., $\hat{y}=G(x)$, minimizing the distance between $\hat{y}$ and $y$.
 
\begin{figure*}[h!]
    \includegraphics[width=\textwidth]{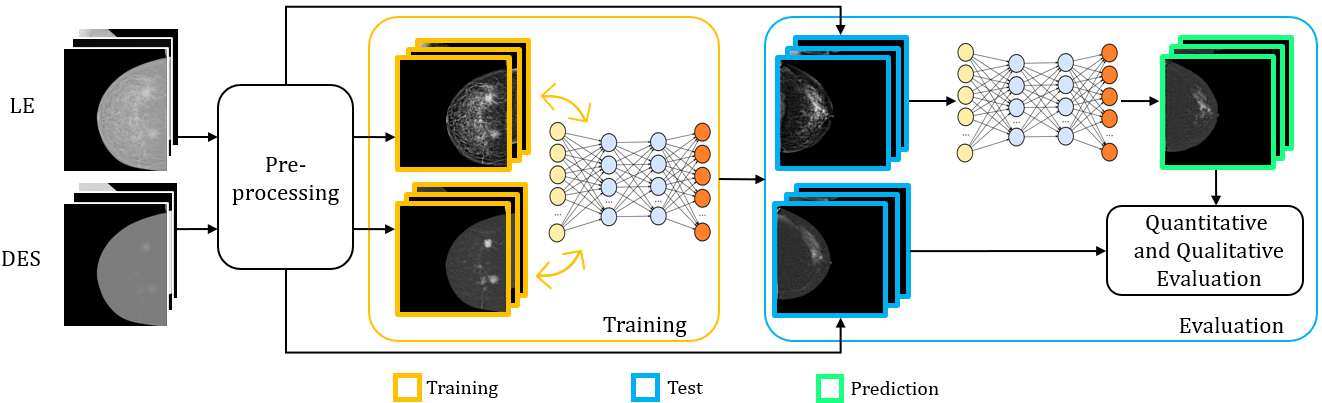}
    \caption{Schematic representation of the methodology.}
    \label{fig:study}
\end{figure*}

A schematic representation of the methods is shown in \figurename~\ref{fig:study} and, accordingly, in this Section we present the image pre-processing, the AI models, and the training and evaluation process.

\subsection{Pre-processing}  \label{subsec:preProcessing}
To ensure that data are consistent and homogeneous, we first applied zero-padding to make images squared, second we applied contrast stretching to adjust the brightness of the images, third we normalized the pixel values bringing them in the range $[0, 1]$, and fourth we resized the images to $256 \times 256$.

\subsection{AI Models}
\begin{figure*}[h!] 
    \includegraphics[width=\textwidth]{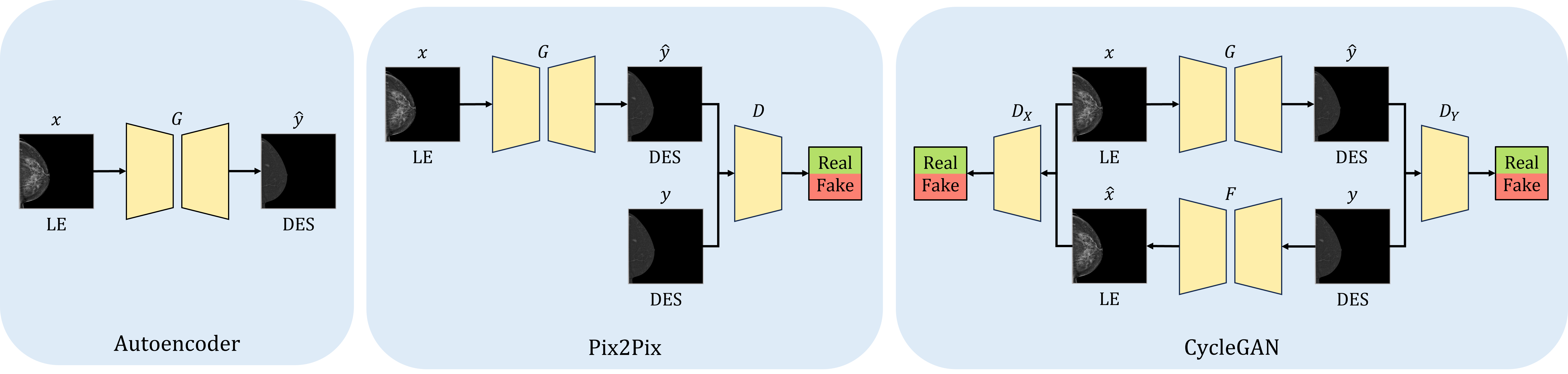}
    \caption{Representative diagrams of Autoencoder, Pix2Pix and CycleGAN translating from LE to DES images. Simbols: $x$ is the input image, $\hat{y}$ is the output image, $y$ is the target image, $G$ and $F$ are the generators, $D$ are the discriminators.}
    \label{fig:architectures}
\end{figure*}

The AI models used in this work to perform the VCE task on CESM images are an autoencoder and two GANs, the Pix2Pix~\cite{pix2pix} and the CycleGAN~\cite{cyclegan}, since these networks have accelerated the use of deep neural networks in image-to-image translation~\cite{overview} and are consistent with the state of the art of VCE, as reported in Section~\ref{sec:introduction}.
Representative diagrams of the architectures used is in \figurename~\ref{fig:architectures}. 

\subsubsection{Autoencoder}
The Autoencoder architecture $G$ consists of four convolutional encoder blocks and four convolutional decoder blocks, with bypass connection and residual connection.
Each block is equipped with three convolutional layers with $3 \times 3$ kernels, each followed by batch normalization, a ReLu activation function and a batch normalization. Only the last decoder block is an exception, as it consists of a convolutional layer with a $1 \times 1$ kernel, followed by a sigmoid activation function.
In the first half of the network, each block is followed by a max pooling layer, while an upsampling layer is used in the later half of the network.
The encoder part takes as input an image $x$ and compresses it into a lower-dimension internal representation.
The decoder part decodes the compact representation provided by the encoder and generates the output image $\hat{y}$.
The loss function minimizes the difference between the generated image $\hat{y}=G(x)$ and the corresponding target image $y$, enabling the optimization of the network parameters as follows:
\begin{equation}
   G^* = \text{arg}\min_{G}\mathbb{E}_{x,y}[||y-G(x)||]
   \label{eq:ae_objective}
\end{equation}
where G* is the optimal model that minimizes the pixel-wise difference denoted as $||.||$ . 

\subsubsection{Pix2Pix}
The Pix2Pix~\cite{pix2pix} is a conditional GAN specifically designed for image-to-image translation with paired datasets.
It comprises a generator network $G$ and a discriminator network $D$, as in a conventional GAN.
The generator $G$ learns the mapping from the $X$ domain to the $Y$ domain.
The $L1$ loss $\mathcal{L}_{L1}(G)$ is used to minimize the difference between the generated output $G(x)~=~\hat{y}$ and the real image $y$, such that $G(x)~=~\hat{y}~\approx y$. It is formulated as:
\begin{equation}
   \mathcal{L}_{L1}(G) = \mathbb{E}_{x,y}[||y-G(x)||_1]
   \label{eq:pix2pix_l1}
\end{equation}
On the other hand, the discriminator $D$ tries to distinguish between the real image $y$ and the generated output $\hat{y}$. 
The generator and discriminator are thus trained in a competitive way using an adversarial loss $\mathcal{L}_{cGAN}(G,D)$, given by:
\begin{equation}
   \mathcal{L}_{cGAN}(G,D) = \mathbb{E}_{x,y}[\log(D(x,y))]+\mathbb{E}_{x}[\log(1-D(x,G(x))]
   \label{eq:pix2pix_lgan}
\end{equation}
The generator tries to minimize this function, learning to fool the discriminator, and the discriminator tries to maximize this function, learning to distinguish real images from synthetic images.
The Pix2Pix full objective is formulated as
\begin{equation}
   G^* = \text{arg}\min_{G}\max_{D}[\mathcal{L}_{cGAN}(G,D) + \lambda\mathcal{L}_{L1}(G)]
   \label{eq:pix2pix_objective}
\end{equation}
where $\lambda$ balances the relative contribution of the two loss terms.
As presented in~\cite{pix2pix}, both the generator network and discriminator network use modules of the form convolution, batch normalization and ReLu.
The generator architecture is based on the U-Net autoencoder~\cite{pix2pix}, with skip connections concatenating encoder block $i$ and decoder block $n-i$, where $n$ is the total number of blocks. 
The discriminator architecture is a convolutional PatchGAN classifier, that tries to classify if each $70 \times 70$ patch in an image is real or fake. By averaging over all image's patches, the overall response of the discriminator is obtained.

\subsubsection{CycleGAN}
The CycleGAN is a type of GAN that learns to translate images from one domain to another without the need for paired samples between the two domains~\cite{cyclegan}.
It comprises two generators, $G$ and $F$, and two discriminators, $D_y$ and $D_x$.
The generator $G$ learns the mapping from the $X$ domain to the $Y$ domain, while the generator $F$ learns the mapping from the $Y$ domain to the $X$ domain.
The discriminator $D_y$ tries to distinguish between the real image $y$ in the $Y$ domain and the generated image $\hat{y~=~G(x)}$.
In the same way, the discriminator $D_x$ tries to distinguish between the real image $x$ in the $X$ domain and the generated image $\hat{x}~=~F(y)$.
To minimize the difference between the distributions of real and generated images, the adversarial loss $\mathcal{L}_{GAN}(G,D_y)$ is used to adversarially train the generator $G$ and the discriminator $D_y$, with $G$ trying to maximize $\mathcal{L}_{adv}(G,D_y)$ to fool $D_y$, while $D_y$ tries to minimize it. It is expressed as:
\begin{equation}
   \mathcal{L}_{GAN}(G,D_Y) = \mathbb{E}_{y}[\log(D_Y(y))]+\mathbb{E}_{x}[\log(1-D_Y(G(x))]
   \label{eq:cg_lgan}
\end{equation}

Similarly, the adversarial loss $\mathcal{L}_{GAN}(F,D_x)$ is used to train adversarially the generator $F$ and the discriminator $D_x$.
To have consistent mappings, the generators should satisfy the forward cycle consistency, being able to bring $x$ back to the original image, i.e. $x~\rightarrow~G(x)~\rightarrow~F(G(x))~\approx~x$, and the backward cycle consistency, that is $y~\rightarrow~F(y)~\rightarrow~G(F(y))~\approx~y$.
This behavior is encouraged using a cycle consistency loss $\mathcal{L}_{cyc}(G, F)$, that helps minimize the difference between $x$ and $F(G(x))$, and between $y$ and $G(F(y))$. It is formulated as:
\begin{equation}
   \mathcal{L}_{cyc}(G,F) = \mathbb{E}_{y}[||G(F(y)-y||_1]+\mathbb{E}_{x}[||F(G(x))-x||_1]
   \label{eq:cg_lcyc_1}
\end{equation}
Moreover, an identity mapping loss $\mathcal{L}_{id}(G,F)$ can be used to ensure that when real samples of the target domain are provided as input to the generator, they are identically mapped to the output. It is formulated as:
\begin{equation}
   \mathcal{L}_{id}(G,F) = \mathbb{E}_{y}[||G(y)-y||_1]+\mathbb{E}_{x}[||F(x)-x||_1]
   \label{eq:cg_lcyc_2}
\end{equation}
The CycleGAN full objective can thus be expressed as:
\begin{equation}
   G^*, F^* = \text{arg} \min_{G, F}\max_{D_x, D_y}[\mathcal{L}_{GAN}(G, D_y) + \mathcal{L}_{GAN}(F, D_x) + \lambda_1\mathcal{L}_{cyc}(G, F) + \lambda_2\mathcal{L}_{id}(G, F)]
   \label{eq:cg_objective}
\end{equation}
where $\lambda_1$ and $\lambda_2$ balance the relative contribution of the loss terms.

The architecture of the generators consists of 3 convolutional layers for down-sampling, multiple residual blocks, 2 transposed convolutional layers for up-sampling, and a final convolutional layer. On the other hand, the architecture of the discriminators is that of a PatchGAN, examining $70 \times 70$ patches to classify the image as real or synthetic. The overall architecture is in line with the one presented in~\cite{cyclegan}.

\subsection{Training}
The Autoencoder, Pix2Pix and CycleGAN models were trained and evaluated on the pre-processed CESM@UCBM dataset, so that they take LE images $x$ as input and generate the virtually contrast-enhanced DES images $\hat{y}$ as output, as similar as possible to the corresponding real DES images $y$. 
All the experiments are computed in 10-fold cross-validation, with the training, validation, and test sets following the respective proportions: 0.8, 0.1, and 0.1.
To facilitate the training, a pre-training on the publicly available dataset of CESM images~\cite{PublicDataset} was performed, following the same pre-processing as described in Section~\ref{subsec:preProcessing}.

To prevent overfitting, we proposed a random data augmentation on the images of the training set in terms of: vertical or horizontal shift (by a maximum of $\pm$10\% of the original dimension), zoom (by a maximum of $\pm$10\%), horizontal flip, and rotation (by a maximum of $\pm$15°).
Each model is trained for a maximum of 200 epochs using an early stopping criterion of 50 epochs following the validation loss.

To train the Autoencoder we used the $L1$ loss and we choose the Adam optimizer for optimizing the model, with a learning rate of $10^{-3}$, a weight decay of $10^{-5}$, a beta of $0.9$ and a momentum of $1$. 

For the Pix2Pix, the generator network $G$ and the discriminator network $D$ were trained simultaneously. 
Referring to the equation \ref{eq:pix2pix_objective}, the binary cross entropy was used as the adversarial loss $\mathcal{L}_{GAN}(G,D)$, and the $L1$ loss $\mathcal{L}_{L1}(G)$ with a $\lambda$ factor of 100 was added for the generator's training. 
For both the generator network and the discriminator network, the Adam optimizer is used with a learning rate of $2\cdot10^{-4}$, a weight decay of $10^{-5}$, a beta of $0.5$ and a momentum of $1$. 

For the CycleGAN, the two generator networks, $G$ and $F$, and the two discriminator networks, $D_y$ and $D_x$, are trained simultaneously.
Typically, the CycleGAN can work with unpaired datasets, but in this work, having access to a paired dataset, the image pairing was maintained in the loss computation and the virtually produced DES images were compared to the corresponding real DES image.
Referring to the equation \ref{eq:cg_objective}, the mean squared error was used for the adversarial losses $\mathcal{L}_{GAN}(G,D_y)$ and $\mathcal{L}_{GAN}(F,D_x)$, the $L1$ loss was both used as the cycle consistency loss with a $\lambda_1$ factor of 10 and as the identity mapping loss with a $\lambda_2$ factor of 5.
For both generator and discriminator networks, the Adam optimizer is used with a learning rate of $10^{-5}$, a weight decay of $10^{-5}$, a beta of $0.5$ and a momentum of $1$. 

For all the models, we do not further investigate any other hyperparameters configuration, since their tuning is out of the scope of this manuscript.
Nevertheless, the “No Free Lunch" Theorem for optimization states that there is no universal set of hyperparameters that will optimize the performance of a model across all possible datasets. 


\subsection{Quantitative Evaluation}

The quantitative analysis was performed by calculating four metrics between the synthetic DES images $\hat{y}$ and the target DES images $y$. 
They are: mean squared error (MSE), peak-signal-to-noise ratio (PSNR), structural similarity index (SSIM) and visual information fidelity (VIF).

The MSE measures the mean squared difference between the pixel values of the target image $y$ and the reconstructed image $\hat{y}$, thus is formulated as:
\begin{equation}
    \text{MSE}(y,\hat{y}) = \frac{1}{mn}\sum_{i=1}^{m}\sum_{j=1}^{n}(y_{ij}-\hat{y}_{ij})^2
   \label{eq: mse}
\end{equation}
where $m$ and $n$ are the number of rows and columns in the images, respectively, and $y_{ij}$ and $\hat{y}_{ij}$ represent the pixels elements at the
$i$th row and $j$th column of $y$ and $\hat{y}$, respectively.
It varies in the range $[0, \infty]$; the lower its value, the higher the quality of the reconstructed image.

The PSNR is defined as the ratio of the maximum possible power of a signal to the power of the noise corrupting the signal.
In our case, the signal is the target image, while the noise is the error introduced by its reconstruction.
The PSNR is typically expressed in decibels (dB), where a PSNR value of 30 dB is considered excellent quality, a value of 27 dB is considered good quality, a value of 24 dB is considered poor quality, and a value of 21 dB is considered bad quality~\cite{psnr_quality}.
It is commonly expressed as a function of the MSE as follows:
\begin{equation}\begin{split}
    \text{PSNR}(y, \hat{y}) = 10 \cdot \log_{10} \left( \frac{\text{max}^2(y)}{\text{MSE}(y, \hat{y})} \right)
   \label{eq: psnr}
\end{split}\end{equation}

The VIF~\cite{VIF} evaluates the quality of visual information in the synthetic image $\hat{y}$ compared to the target image $y$. It is derived by a modeling of the Human-Visual-System in the wavelet domain and is formulated as:
\begin{equation}\begin{split}
    \text{VIF}(y, \hat{y}) = \frac{\sum_{j\in{subbands}}I(y^{j})}{\sum_{j\in{subbands}}I(\hat{y}^{j})}
    \label{eq: vif}
\end{split}\end{equation}
where $I(y^{j})$ and $I(\hat{y}^{j})$ represent the information ideally extracted by the brain from a certain subband in the target image and synthetic image, respectively.
It usually varies in the range $[0, 1]$, the higher its value, the higher the quality of the synthetic image. A VIF $> 1$ indicates that the synthetic image has superior quality to the target image~\cite{VIF}.

The SSIM~\cite{ssim} measures the similarity between two images by comparing luminance, contrast and structure. The formulation is as follows:
\begin{equation}
    \text{SSIM}(y, \hat{y}) = \frac{{(2\mu_y\mu_{\hat{y}} + C_1)(2\sigma_{y\hat{y}} + C_2)}}{{(\mu_y^2 + \mu_{\hat{y}}^2 + C_1)(\sigma_y^2 + \sigma_{\hat{y}}^2 + C_2)}}
    \label{eq:ssim}
\end{equation}
where $\mu_y$ and $\mu_{\hat{y}}$ represent the means of the images $y$ and $\hat{y}$ respectively, while $\sigma_y$ and $\sigma_{\hat{y}}$ represent their standard deviations. $C_1$ and $C_2$ are small constants used for stabilization.
The SSIM varies in the range $[0, 1]$, the higher its value, the greater the similarity between the two images~\cite{ssim}.

\subsection{Qualitative Evaluation}
The goal of the  qualitative analysis is to have a medical assessment on the synthetic DES images generated by the models.
It consists of  two visual Turing tests presented to four expert radiologists with experience ranging from 8 to 38 years.
For both assessments, we set up a replicated radiology reading room environment, and we randomly selected images from 50 patients of the CESM@UCBM test sets from the different cross-validation folds.
To have a global consensus, the final human-based decisions were given by majority voting  weighted by the years of experience of the radiologists~\cite{voting}.

The first visual Turing test helps find  the DL model producing the most realistic synthetic DES images.
To this end, we showed radiologists side-by-side synthetic DES images generated by the Autoencoder, Pix2Pix and CycleGAN for the same LE input image.
Then, each radiologist had to independently choose the most realistic image, without knowing which model generated them. 

The second visual Turing test aims to study not only if synthetic images are perceived as real ones but also if they can be used to assign the BI-RADS score, which is a measure indicating patients' risk of developing breast cancer (Section~\ref{sec:materials}).
We set the input size of the  model winning the first visual Turing test to $512 \times 512$, and then we retrained it. 
Note that this is a higher image resolution  than the one used in previous experiments, i.e.,  $256 \times 256$ as reported in Section~\ref{subsec:preProcessing}: while this has the beneficial effect of producing more realistic images, it substantially increases the computational costs.
Hence, we     provided the four radiologists with both real DES images and these newly generated synthetic DES images. 
All the images are presented one at a time, in random order: the radiologist specified whether the displayed image is real or synthetic, and he/she scored  the BI-RADS.

\section{Results and Discussion} \label{sec:res and dis}

\begin{figure*}[h!] 
    \includegraphics[width=\textwidth]{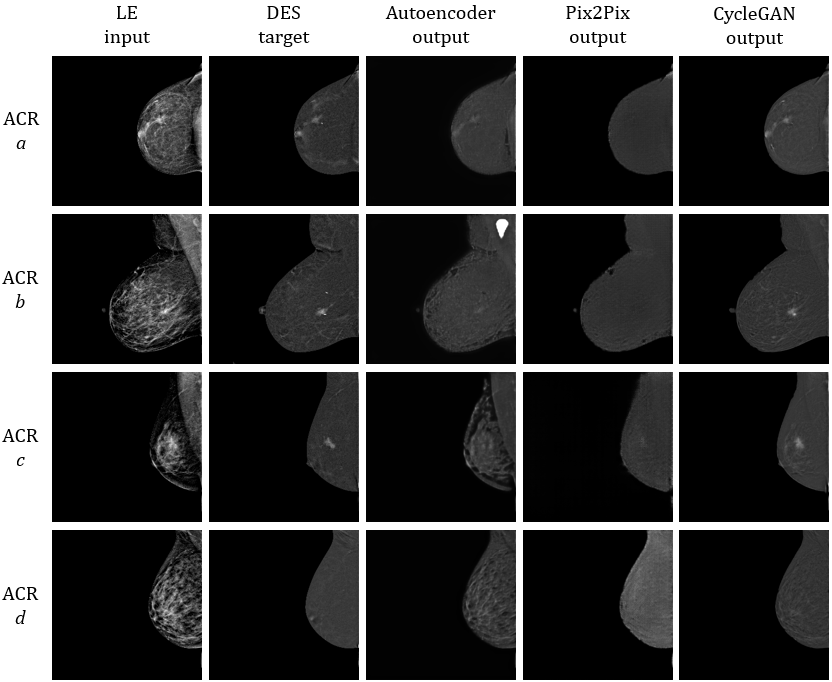}
    \caption{Four representative cases of LE input images, DES target images, and output images of Autoencoder, Pix2Pix, and CycleGAN (from left to right) varying the ACR category (from top to bottom).}
    \label{fig:imm}
\end{figure*}

\tablename~\ref{tab1} shows the outcomes of the quantitative analysis performed for the Autoencoder, Pix2Pix and CycleGAN on the whole test sets.
For each model, the metrics MSE, PSNR, VIF, and SSIM are calculated between all the target DES images of the test sets from the 10 folds and their corresponding synthetic DES images.
The metrics' values are reported as the mean value $\pm$ standard deviation calculated on the different folds.
For each metric, the most performing model is highlighted in bold.
The Autoencoder obtained the worst mean value in the case of MSE, PSNR and SSIM.
As shown in \figurename~\ref{fig:imm}, it produces images that are generally blurry, often too similar to the input image rather than the corresponding target image, or characterized by the presence of non-physiological hyperintense areas.
These results are in line with expectations, as autoencoders were among the first methods used for image-to-image translation, and it is known that they have the limitation of generating blurry images, partly due to the element-wise criteria typically used, such as $L1$ loss and $L2$ loss~\cite{limiteae1, limiteae2}.
The Pix2Pix obtains the worst mean value of VIF but also the best mean value of SSIM, while the CycleGAN obtains the best mean values of MSE, PSNR and VIF. 
As displayed in \figurename~\ref{fig:imm}, compared to the Autoencoder, Pix2Pix and CycleGAN produce output images that are more similar to the target.
However, the images produced by the Pix2Pix do not have sufficient informative content, as proven by the low VIF value.
The Pix2Pix often generates images where the virtual enhancement of contrast corresponding to the lesion is missing and the obtained texture is different from the target.
On the contrary, as suggested by the metrics and evident from \figurename~\ref{fig:imm}, the CycleGAN proves to be the most promising network in generating synthetic DES images, as it is the one that best manages to reconstruct the contrast enhancement.

It is worth noting that such results agree with those reported by Azarfar et al, who implemented VCE on CT~\cite{reviewvcect}: indeed, they proved that the CycleGAN outperforms conditional GANs in terms of performance, likely due to its ability to learn heuristically from all the images in the paired dataset without being restricted to specific pixel relationships as the Pix2Pix.

\begin{table*}[h] 
\begin{center}
\resizebox{\columnwidth}{!}{
\begin{tabular}{ccccc}
\toprule
\textbf{Model} & \textbf{MSE ↓} & \textbf{PSNR ↑} & \textbf{VIF ↑} & \textbf{SSIM ↑} \\ 
\midrule
Autoencoder & 0.0083 ± 0.0036 & 23.2775 ± 2.4942 & 0.1581 ± 0.0212 & 0.5365 ± 0.2104 \\ 
Pix2Pix     & 0.0042 ± 0.0008 & 26.3476 ± 0.7640 & 0.1129 ± 0.0106 & \textbf{0.8575 ± 0.0132} \\ 
CycleGAN    & \textbf{0.0038 ± 0.0010} & \textbf{26.4866 ± 0.9206} & \textbf{0.1840 ± 0.0132} & 0.8492 ± 0.0131 \\ 
\bottomrule
\end{tabular}}
\caption{Quantitative analysis performed for Autoencoder, Pix2Pix and CycleGAN on the whole test set. Each tabular denotes the mean $\pm$ standard deviation from the 10-folds.}
\label{tab1}
\end{center}
\end{table*}

After obtaining an overall measure of the models' performance by calculating metrics on the entire test sets, it is important to verify that the models are robust to different mammographic densities and that their performance does not deteriorate in the most critical cases, represented by the ACR categories \textit{c} (heterogeneously dense breast tissue) and \textit{d} (extremely dense breasts).
Indeed, heterogeneously or extremely dense breast parenchyma can cause tumor lesions to not be clearly identifiable in images generated by FFDM.
Therefore, in these cases, there is a risk that DL models my find it more complex to generate virtual DES images using only LE images, which are analogous to FFDM images.
For these reasons, we also compute the MSE, PSNR, VIF, and SSIM for each model by distinguishing the test sets' images according to the breasts' ACR categories \textit{a}, \textit{b}, \textit{c}, and \textit{d} which define mammographic density.
\tablename~\ref{tab:acr} reports the outcomes of this analysis, with the metrics values reported as mean value $\pm$ standard deviation calculated on the different folds.
Within each model, the values of a specific metric are compared across the different categories with the best values highlighted in bold.
However, our analysis indicates that there are no statistically significant differences (Kruskal-Wallis $p$-value $>$ 0.05) in the values of the same metric when calculated for the four categories within each model.
Additionally, it is worth noting that performance does not deteriorate in the critical cases \textit{c} and \textit{d}.
Hence, these findings suggest that the implemented models demonstrate robustness across the different mammographic densities.

\begin{table*}[h] 
\begin{center}
\resizebox{\columnwidth}{!}{
\begin{tabular}{cccccc}
\toprule
\multicolumn{1}{c}{\textbf{Model}} & \multicolumn{1}{c}{\textbf{ACR}} &\multicolumn{1}{c}{\textbf{MSE ↓}} & \multicolumn{1}{c}{\textbf{PSNR ↑}} & \multicolumn{1}{c}{\textbf{VIF ↑}} & \textbf{SSIM ↑} \\ 
\midrule
\multicolumn{1}{c}{\multirow{4}{*}{Autoencoder}} & \textit{a} & \multicolumn{1}{c}{0.0108 $\pm$ 0.0048} & \multicolumn{1}{c}{21.6093 $\pm$ 2.9097} & \multicolumn{1}{c}{0.1384 $\pm$ 0.0270} & 0.5054 $\pm$ 0.2055 \\ 
\multicolumn{1}{c}{} & \textit{b} & \multicolumn{1}{l}{0.0114 $\pm$ 0.0052} & \multicolumn{1}{l}{21.7321 $\pm$ 2.9283} & \multicolumn{1}{l}{\textbf{0.1676 $\pm$ 0.0249}} & \textbf{0.5426 $\pm$ 0.2071} \\ 
\multicolumn{1}{l}{} & \textit{c} & \multicolumn{1}{l}{0.0065 $\pm$ 0.0031} & \multicolumn{1}{l}{23.8066 $\pm$ 2.2557} & \multicolumn{1}{l}{0.1566 $\pm$ 0.0247} & 0.5358 $\pm$ 0.2110 \\ 
\multicolumn{1}{l}{} & \textit{d} & \multicolumn{1}{l}{\textbf{0.0053 $\pm$ 0.0024}} & \multicolumn{1}{l}{\textbf{24.5332 $\pm$ 2.7948}} & \multicolumn{1}{l}{0.1612 $\pm$ 0.0256} & 0.5274 $\pm$ 0.2108 \\ 
\midrule
\multicolumn{1}{c}{\multirow{4}{*}{Pix2Pix}} & \textit{a} & \multicolumn{1}{l}{0.0034 $\pm$ 0.0014} & \multicolumn{1}{l}{25.8957 $\pm$ 1.4328} & \multicolumn{1}{l}{0.1505 $\pm$ 0.0322} & 0.8170 $\pm$ 0.0529 \\ 
\multicolumn{1}{l}{} & \textit{b} & \multicolumn{1}{l}{0.0047 $\pm$ 0.0026} & \multicolumn{1}{l}{26.0185 $\pm$ 2.3233} & \multicolumn{1}{l}{0.1807 $\pm$ 0.0173} & 0.8365 $\pm$ 0.0207 \\ 
\multicolumn{1}{l}{} & \textit{c} & \multicolumn{1}{l}{\textbf{0.0029 $\pm$ 0.0009}} & \multicolumn{1}{l}{\textbf{27.1512 $\pm$ 1.4586}} & \multicolumn{1}{l}{0.1738 $\pm$ 0.0139} & \textbf{0.8546 $\pm$ 0.0129} \\ 
\multicolumn{1}{l}{} & \textit{d} & \multicolumn{1}{l}{0.0031 $\pm$ 0.0013} & \multicolumn{1}{l}{27.0355 $\pm$ 1.2134} & \multicolumn{1}{l}{\textbf{0.1917 $\pm$ 0.0310}} & 0.8509 $\pm$ 0.0424 \\ 
\midrule
\multicolumn{1}{c}{\multirow{4}{*}{CycleGAN}} & \textit{a} & \multicolumn{1}{l}{0.0055 $\pm$ 0.0036} & \multicolumn{1}{l}{24.7935 $\pm$ 1.8855} & \multicolumn{1}{l}{0.1582 $\pm$ 0.0320} & 0.8189 $\pm$ 0.0500 \\ 
\multicolumn{1}{l}{} & \textit{b} & \multicolumn{1}{l}{0.0060 $\pm$ 0.0038} & \multicolumn{1}{l}{25.9411 $\pm$ 2.0808} & \multicolumn{1}{l}{\textbf{0.1982 $\pm$ 0.0147}} & 0.8486 $\pm$ 0.0196 \\ 
\multicolumn{1}{l}{} & \textit{c} & \multicolumn{1}{l}{0.0039 $\pm$ 0.0016} & \multicolumn{1}{l}{26.2804 $\pm$ 1.1737} & \multicolumn{1}{l}{0.1827 $\pm$ 0.0135} & \textbf{0.8564 $\pm$ 0.0103} \\ 
\multicolumn{1}{l}{} & \textit{d} & \multicolumn{1}{l}{\textbf{0.0028 $\pm$ 0.0007}} & \multicolumn{1}{l}{\textbf{26.9839 $\pm$ 1.1798}} & \multicolumn{1}{l}{0.1859 $\pm$ 0.0332} & 0.8439 $\pm$ 0.0497 \\ 
\bottomrule
\end{tabular}}
\caption{Quantitative analysis performed for Autoencoder, Pix2Pix and CycleGAN distinguishing the test set images according to the breast's ACR categories \textit{a}, \textit{b}, \textit{c}, and \textit{d}.}
\label{tab:acr}
\end{center}
\end{table*}

\begin{figure*}[h!] 
\centering
    \includegraphics[scale=0.5]{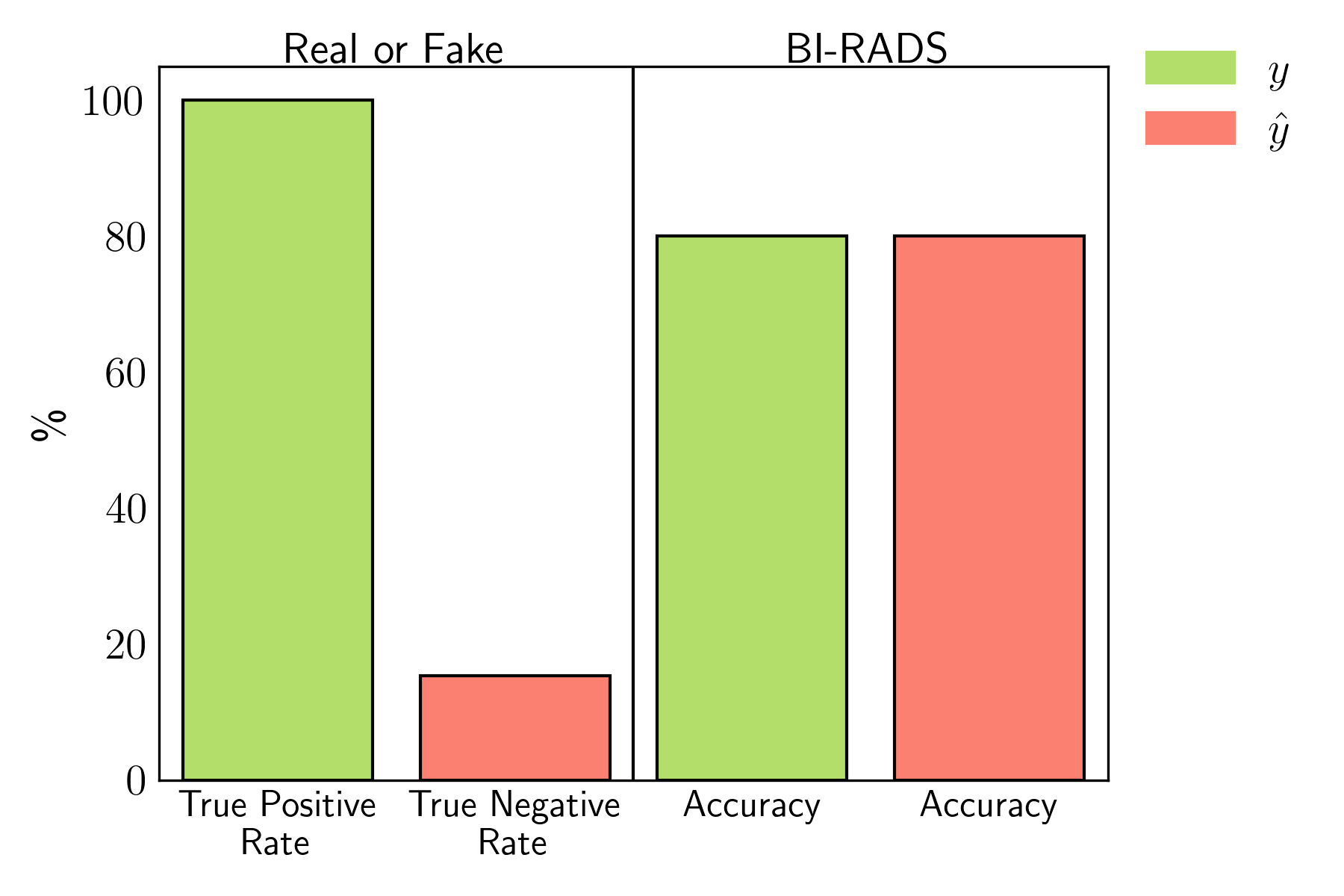}
    \caption{Results of the second visual Turing test. True Positive Rate and True Negative Rate are the percentages at which the radiologists correctly label real images $y$ and synthetic images $\hat{y}$, respectively. 
    The Accuracy is the percentage at which the radiologists correctly assign the BI-RADS score to $y$ and $\hat{y}$.}
    \label{fig:graf}
\end{figure*}

In addition to the quantitative analysis, also the qualitative analysis carried out with the support of radiologists has been crucial in evaluating the performance of the DL models.
The first visual Turing test was proposed to identify the model capable of generating the most realistic DES images.
In this regard, for each set of images shown, all radiologists consistently determined that the images most realistic were those generated by the CycleGAN, choosing it as the most reliable network for all the cases. This is in agreement with the results obtained from the quantitative analysis.
For this reason, we further investigated the CycleGAN, and we conducted the second visual Turing test by using both the real DES images and the CycleGAN-generated DES images.
The results are graphically represented in the \figurename~\ref{fig:graf}, which shows the True Positive Rate and True Negative Rate obtained by radiologists in distinguishing real images from synthetic images, as well as the Accuracy in associating the BI-RADS descriptor on real and synthetic images.
In order to calculate the True Positive Rate and True Negative Rate, we have assigned positive samples to real DES images $y$ and negative samples to synthetic DES images $\hat{y}$.
Therefore, it is desirable to achieve a high True Positive Rate and a low True Negative Rate.
The True Positive Rate is higher when radiologists are better at labelling real DES images, reflecting their reliability.
The True Negative Rate is higher when radiologists are better at labelling synthetic DES images.
Consequently, a low True Negative Rate indicates that the network generates realistic synthetic DES images that can confuse radiologists into thinking they are real.

On the other hand, we evaluated the Accuracy in assigning the BI-RADS descriptor separately for real DES images and synthetic DES images. In both subsets, it was calculated as the ratio of correctly assigned descriptors to the total number of descriptors assigned.
Therefore, obtaining a high Accuracy value is desirable for both real and synthetic images.

The results show that a True Positive Rate of 100\% has been achieved, meaning that all true DES images were correctly recognized, and a True Negative Rate of 15.4\%, indicating that only 15.4\% of the synthetic DES images were identified as fake, while the remaining 84.6\% were classified as real.

Additionally, the Accuracy in assigning the BI-RADS descriptor was found to be 80\% for both real DES images and synthetic DES images. Thus, the percentage of error on real DES images is the same as the error on synthetic DES images.

These results suggest that the synthetic DES images have the potential to be used for diagnosis, so it is possible to apply the VCE on CESM images by using AI techniques.
Thereby, the advantages of CESM, such as better diagnostic accuracy compared to FFDM, especially in patients with dense breast tissue, could be achieved without the requirement of intravenous administration of contrast medium or the need for dual-energy acquisition to obtain the HE image alongside the LE image.

It is worth noting, however, that although the proposed method assumes no administration of contrast agent, the LE images, used for the present study, were actually acquired after the administration of the contrast agent.
Nevertheless, the assumption that the LE images are contrast-free images is justified because the LE images were acquired in an energy range where the influence of iodine k-edge is considered to be negligible. 

\section{Conclusions} \label{sec:conclusion}
CESM is a breast imaging technique that produces LE images, similar to FFDM, and DES images where a tumor lesion can be easily seen thanks to an injected iodine contrast medium.
In this context, our work offers two contributions.
First, this study has  demonstrated that VCE  can be applied to  CESM images avoiding    the need for contrast medium administration and minimizing radiation exposure. 
This claim is supported by experiments using three well-established generative deep models, namely an autoencoder, Pix2Pix, and CycleGAN, that  generate DES images solely from LE images.
We found that CycleGAN provides the best-quality synthetic images:  in many  cases they are hard to be distinguished from real images even for a pool of four  radiologists.
Furthermore, the results suggest  that BI-RADS scores can be assigned from CycleGAN-generated images with an accuracy equal to using real images. 

Second, we have presented and made publicly accessible a unique dataset of CESM images. 
We have taken great care to ensure the dataset's value to the scientific community by including  data extracted from the medical  reports and from the  biopsy, and  also the DICOM images. 
Indeed this offers image information available in the DICOM tags, which  would not be accessible in JPEG format as is the case of the other CESM public dataset~\cite{PublicDataset}.
We hope that our   dataset provides all the necessary information for conducting CESM analyses, and we extend an invitation to the scientific community to delve deeper into this area.

Despite these promising findings, the way to the clinical adoption of VCE technology in clinical practice needs further research and  validation,  including the development of custom deep architectures and the incorporation of loss functions that prioritize lesion reconstruction during training.
We are working to enlarge the dataset and to set up external validation involving other centers. 
Furthermore, while here we have  generated full-dose DES images from LE images that simulate images with zero-dose of contrast medium, we deem that it would be interesting to implement models  producing full-dose DES images from low-dose DES images as this approach holds the potential for even more improved outcomes.

\section*{CRediT authorship contribution statement}
\textbf{Aurora Rofena}: Conceptualization, Methodology, Software, Validation, Formal analysis, Investigation, Data Curation, Writing - Original Draft, Writing - Review \& Editing, Visualization;
\textbf{Valerio Guarrasi}: Conceptualization, Methodology, Software, Validation, Formal analysis, Investigation, Data Curation, Writing - Original Draft, Writing - Review \& Editing, Visualization, Supervision, Project administration;
\textbf{Marina Sarli}: Resources, Data Curation;
\textbf{Claudia Piccolo}: Conceptualization, Resources;
\textbf{Matteo Sammarra}: Conceptualization, Resources;
\textbf{Bruno Beomonte Zobel}: Conceptualization, Resources;
\textbf{Paolo Soda}: Conceptualization, Methodology, Writing - Review \& Editing, Supervision, Project administration, Funding acquisition.

\section*{Conflict of interest statement}
All authors declare that they have no known competing financial interests or personal relationships that could have appeared to influence (bias) the work reported in this paper.

\section*{Acknowledgements}

Aurora Rofena is a Ph.D. student enrolled in the National Ph.D. in Artificial Intelligence, XXXVIII cycle, course on Health and life sciences, organized by Università Campus Bio-Medico di Roma.

This work was partially founded by the project n. F/130096/01-05/X38 -  Fondo per la Crescita Sostenibile - ACCORDI PER L'INNOVAZIONE DI CUI AL D.M. 24 MAGGIO 2017 - Ministero dello Sviluppo Economico (Italy), and from PNRR MUR project PE0000013-FAIR.

Resources are provided by the National Academic Infrastructure for Supercomputing in Sweden (NAISS) and the Swedish National Infrastructure for Computing (SNIC) at Alvis @ C3SE, partially funded by the Swedish Research Council through grant agreements no. 2022-06725 and no. 2018-05973.

\bibliographystyle{unsrtnat}
\bibliography{mybibfile}

\end{document}